\documentclass[review]{elsarticle}
\usepackage{lineno,hyperref}

\usepackage{pgfplots}
\pgfplotsset{compat=newest}
\pgfplotsset{every axis legend/.append style={
cells={anchor=west}}
}
\usepgfplotslibrary{polar,groupplots}
\usetikzlibrary{positioning}
\usetikzlibrary{external}

\usetikzlibrary{arrows}
\tikzset{>=stealth'}

\usepackage{graphicx}
\usepackage{amsmath}
\usepackage{amsfonts}
\usepackage{amssymb}
\usepackage{placeins}
\modulolinenumbers[1]

\journal{Chemical Physics}

\biboptions{sort&compress}

\newcommand{\mean}[1]{\left\langle #1 \right\rangle}

\newcommand{\Tr}{\text{Tr}\,}

\renewcommand{\Re}{\text{Re}\,}
\newcommand{\im}{\mathrm{i}}

\newcommand{\eps}{\varepsilon}

\newcommand{\mbf}[1]{\ensuremath{\mbox{\boldmath$#1$}}}

\newcommand{\RDM}{\ensuremath{\sigma}}

\newcommand{\COMM}[2]{\left[#1, #2 \right]_-}
\newcommand{\ACOMM}[2]{\left[#1, #2\right]_+}

\newcommand{\bSigma}{\mbf{\Sigma}}
\newcommand{\bGamma}{\mbf{\Gamma}}

\begin{document}

\begin{frontmatter}

\title{Time-dependent framework for energy and charge currents in nanoscale systems}

\author[mymainaddress]{Thomas Lehmann}
\author[mymainaddress,mysecondaryaddress]{Alexander Croy}
\author[mymainaddress,mysecondaryaddress]{Rafael Guti\'{e}rrez}
\author[mymainaddress,mysecondaryaddress,mythirdaddress]{Gianaurelio Cuniberti}

\address[mymainaddress]{Institute for Materials Science and Max Bergmann Center of Biomaterials, TU Dresden, 01062 Dresden, Germany}
\address[mysecondaryaddress]{Dresden Center for Computational Materials Science (DCMS), TU Dresden, 01062 Dresden, Germany}
\address[mythirdaddress]{Center for Advancing Electronics Dresden, TU Dresden, 01062 Dresden, Germany}

\begin{abstract}
The calculation of time-dependent charge and energy currents in nanoscale
systems is a challenging task. Nevertheless it is crucial for gaining a deep
understanding of the relevant processes at the nanoscale. We extend
the auxiliary-mode approach for time-dependent charge transport to allow for the
calculation of energy currents for arbitrary time dependencies. We apply the approach to two illustrative examples,
a single-level system and a benzene ring, demonstrating its usefulness for
a wide range of problems beyond simple toy models, such as molecular devices.
\end{abstract}


\end{frontmatter}

\section{Introduction}
The transport and conversion of energy are of central importance for many biological systems and technological applications. Nanotechnology offers a route to make use of the efficiency and robustness found in nature for energy harvesting and energy transduction in nano-scale devices. To take full advantage of the possibilities provided by molecular junctions~\cite{Remade2004,Remacle2006} or quantum dot arrays~\cite{Remacle2004,Mol2011}, a deep understanding of the involved dynamical processes is necessary. Here, the theoretical description of time-dependent transport of charge and energy can contribute valuable insights and help to find new applications~\cite{doi:10.1063/1.4936182,Klymenko2016}.

In this context, there has been considerable interest in time-resolved energy transport and entropy production in nano-scale devices~\cite{crsi+11,crsi+14,dalo16,luli+14,lumo+16}. The main goal of those studies is the formulation of thermodynamical laws for artificial nano systems, which will help to mimic nature and to realize nanomachines. A central difference to charge transport is the ability of the contact region between device and reservoirs to contribute to the energy balance~\cite{luli+14,lumo+16}. Typically the analysis
is restricted to simplified single-level systems, for which analytical results can be obtained.

In general, the numerical description of dynamical processes in realistic nano-scale devices remains very challenging. There exist several approaches~\cite{kurt05,zhu05a,mold07a,prok08a,croy09a,gaur14,ridl15a} which are mostly based on the (time-dependent) nonequilibrium Green’s function (TDNEGF) formalism~\cite{wing93,jauh94,haug96}.
Among those, the auxiliary-mode approach has proven to be versatile and efficient~\cite{crsa09a,pocr16}, and it has successfully been applied to several transport scenarios~\cite{croy12a,croy12b,pope12a,chen14c,xie13a,cao15a,wang13a,pocr17,kuel+17}. The relevance of time-dependent processes also reaches into the realm of biological systems. As it has been discussed in many studies~\cite{pope12a,PhysRevLett.102.208102,PhysRevB.82.155455,doi:10.1021/ja078162j,doi:10.1021/jp004436c}, structural fluctuations are able to gate electron transport in biomolecular systems, thus requiring a description going beyond purely static transport.

In this contribution, we extend the auxiliary-mode formalism to allow for the calculation of energy currents carried by the transferred electrons. We show that those currents can be obtained without additional effort from known quantities within the wide-band limit (WBL). To demonstrate the new method, we consider two different scenarios. Firstly, we study the response of a single-level device to a nearly rectangular voltage-pulse. We verify that the total charge and energy currents fulfill the expected conservation laws and attain the correct steady-state values for long pulses. Further, we find that both currents show a similar transient behavior. Secondly, we show that the method can also be applied to a more realistic system, thus opening the possibility of combining it with an atomistic description of nanoscale systems. To this purpose, we consider a benzene molecule symmetrically (\textit{para} positions) and asymmetrically (\textit{meta} positions) contacted by one-dimensional electrodes. Asymmetric contact configurations or, alternatively, intrinsic structural asymmetries in molecular-scale junctions, are known to display quantum interference effects in many interesting situations~\cite{ANIE:ANIE201207667,PhysRevB.59.16011,doi:10.1021/acs.jpclett.5b01595,doi:10.1021/acs.jpclett.6b02989,doi:10.1021/nl101688a,doi:10.1063/1.2958275,doi:10.1021/jp9117216}. It is therefore of great interest to elucidate the possible interplay between quantum interference and time-dependent effects. We have recently shown, within a Tien-Gordon approach, that sinus-like AC-fields can be used in planar atomic-scale T-shaped circuits to counteract destructive quantum interference, thus leading to a current increase at low bias~\cite{doi:10.1063/1.4936182}. In this contribution, we use the benzene ring to study the influence of interference on time-dependent charge and energy currents. Overall, the extended method is well suited to investigate energy transport in a variety of nano-scale systems.

The article is organized as follows. In section \ref{sec:model} we present the general model of a driven device coupled to two electronic reservoirs. Subsequently, we summarize the auxiliary-mode approach and then derive general expressions for the energy currents in terms of auxiliary current-matrices. In section \ref{sec:results} we apply the method to the two previously mentioned illustrative examples: a single-level system and a benzene molecule. Finally, section \ref{sec:summary} gives a summary and an outlook.

 \section{Method}\label{sec:model}
\subsection{Model}
We consider the usual threefold setup consisting of a device which is
coupled to two electron reservoirs. The coupling is due to a tunneling between
states in the device and the reservoirs.
The total Hamiltonian is
\begin{equation}
  H = H_\mathrm{dev.}(t) + H_\mathrm{res.} + H_\mathrm{coup.}\,.
  \label{eq:totalHam}
\end{equation}
The device is described in terms of two sets of tight-binding parameters: discrete on-site energies $\varepsilon_n(t)$
and hopping parameters $V_{n m}(t)$,
\begin{equation}\label{eq:DevHamilOp}
  H_\mathrm{dev.} = \sum_n \varepsilon_n(t) c^\dagger_n c_n
  +\sum_{n\ne m} V_{n m}(t) c^\dagger_n c_m \;.
\end{equation}
The operators $\{c^\dagger_n\}$ and $\{c_n\}$ denote the creation and annihilation
of an electron in state $n$. The explicit time-dependence of the device Hamiltonian
can be due to changes of the on-site energies or of the hopping parameters.
The reservoirs are described by non-interacting electrons and the respective Hamiltonian reads
\begin{equation}
  H_\mathrm{res.} = \sum_{\alpha \in \mathrm{L},\mathrm{R}} H_{\alpha} = \sum_{\alpha \in \mathrm{L},\mathrm{R}}
  \sum_k \varepsilon_{\alpha k} b^\dagger_{\alpha k} b_{\alpha k}\;.
  \label{eq:ResHamilOp}
\end{equation}
Initially, before the coupling to the device is switched on, the reservoirs are in thermal
equilibrium characterized by a well-defined temperature $T$ and chemical potential $\mu_\alpha$.
Finally, the coupling Hamiltonian is
\begin{equation}
  H_\mathrm{coup.} = \sum_{\alpha k } H_{c \alpha}= \sum_{\alpha k }\sum_n
      T^\alpha_{k n}(t) \,b^\dagger_{\alpha k} c_n
    + \rm{h.c.} \;,
 \label{eq:TunnHam}
\end{equation}
with $\{T^\alpha_{k n}\}$ denoting the tunnel couplings between device and
reservoir $\alpha={\rm L}, {\rm R}$. The operators $\{b^\dagger_{\alpha k}\}$ and $\{b_{\alpha k}\}$ are
electron creation and annihilation operators for reservoir
states, respectively.

\subsection{Auxiliary-mode approach}
In the following, matrix representations of system operators will be denoted by bold-face
symbols. Throughout the paper we adopt units with $\hbar = 1$.

The single-electron density matrix $\RDM_{mn} = \mean{c^\dagger_n c_m}$ of the device connected
to electronic reservoirs obeys the following equation of motion\cite{zhwa+07}
\begin{equation}
  \im \frac{\partial}{\partial t}\mbf{\RDM}( t )
    = \COMM{ \mathbf{H} (t) }{ \mbf{\RDM}( t ) }
    +\im \sum\limits_{\alpha \in \text{leads}} \left( \mbf{\Pi}_\alpha ( t ) + \mbf{\Pi}^\dagger_\alpha ( t ) \right)\;,
    \label{eq:EOMSEDM1}
\end{equation}
where we have defined so-called current matrices $\mbf{\Pi}_\alpha$. Those can be expressed in terms of Green functions and
self-energies,
\begin{equation}
  \mbf{\Pi}_\alpha ( t ) = \int\limits^t_{t_0} dt_2
     \left( \mathbf{G}^>(t,t_2) \mbf{\Sigma}^<_\alpha (t_2,t)
       - \mathbf{G}^<(t,t_2) \mbf{\Sigma}^>_\alpha (t_2,t) \right) \;,
  \label{eq:DefPi}
\end{equation}
where $\mbf{G}^\gtrless$ and $\mbf{\Sigma}^\gtrless$ are the usual lesser/greater Green functions and self-energies \cite{haug96}, respectively.

In the general case, the calculation of the current matrices is very demanding. In the WBL and for the case where the level-width function $\mbf{\Gamma}_{\alpha}(\eps)$ is approximated by a sum of Lorentzians, an effective propagation scheme has been proposed \cite{crsa09a}.
The key ingredient is the decomposition of the Fermi function $f(\eps)$, which enters the self-energies $\mbf{\Sigma}^\gtrless$,
into a sum of simple poles
\begin{equation}
	f(\varepsilon) = \frac{1}{1 + \exp(\varepsilon)} \approx \frac{1}{2} - \sum_{p=1}^{N_{\rm F}} \left( \frac{R_p}{\varepsilon-z^+_p}+\frac{R_p}{\varepsilon-z^-_p} \right)\;,
	\label{eq:DecFermi}
\end{equation}
where $R_p$ is the $p$th residue and $z_p^{(\pm)}$ is the $p$th pole in the upper $(+)$ and lower $(-)$ complex plane.
The best known decomposition of this type is the Matsubara decomposition for which $R_p=1$ and $z_p^+ = \pi(2p-1)$. A very efficient decomposition is the Pad\'e decomposition \cite{oz07}, for which the residues
and the poles can be efficiently calculated by solving an eigenvalue problem \cite{kame+10}. For the sake of clarity, we restrict the following discussion to the WBL and refer to \cite{crsa09a} for the more general case of a Lorentzian expansion of the level-width functions, leading to a more complex expression for the current matrices.

In the WBL it was found that the current matrices can be expressed by \cite{crsa09a}
\begin{equation}\label{eq:DefPiWBL}
    \mbf{\Pi}_\alpha ( t ) = \frac{1}{4}\left( \mbf{1} - 2\mbf{\RDM}\right)\mbf{\Gamma}_\alpha + \sum_p \mbf{\Pi}_{\alpha p} (t)\;,
\end{equation}
and the auxiliary current-matrices $\mbf{\Pi}_{\alpha p}$ obey
\begin{equation}\label{eq:EOMPi}
    \im \frac{\partial}{\partial t}  \mbf{\Pi}_{\alpha p} = \frac{R_p}{\beta} \mbf{\Gamma}_\alpha
                + \left[ \mbf{H} (t) - \frac{\im}{2} \mbf{\Gamma} - \chi^+_{\alpha p} \mbf{1}\right] \mbf{\Pi}_{\alpha p}\;.
\end{equation}
Here, $\mbf{\Gamma}=\sum_\alpha \mbf{\Gamma}_\alpha$ and $\chi^+_{\alpha p} = \mu_\alpha + \im z_p/\beta$. The inverse temperature $\beta$ is given by $\beta=(k_{{\rm B}}T)^{-1}$. Using Eq.\ \eqref{eq:DefPiWBL} in Eq.\ \eqref{eq:EOMSEDM1} one finally gets
\begin{multline}
  \im \frac{\partial}{\partial t}\mbf{\RDM}( t )
    = \COMM{ \mathbf{H} (t) }{ \mbf{\RDM}( t ) }  - \im \ACOMM{ \mbf{\Gamma}/2 }{ \mbf{\RDM}( t ) } \\
    +\im \sum\limits_{\alpha \in \text{leads}} \left(   \frac{1}{4} \mbf{\Gamma}_\alpha + \sum_p \mbf{\Pi}_{\alpha p} (t) + h.c. \right)\;.
    \label{eq:EOMSEDM2}
\end{multline}
Equations \eqref{eq:EOMPi} and \eqref{eq:EOMSEDM2} form a closed set of first-order differential equations which can be solved for given initial conditions. In practice, the calculations are typically started with $\mbf{\RDM}(t_0)=\mbf{0}$ and $\mbf{\Pi}_{\alpha p} = \mbf{0}$ and the system is equilibrated before any driving is switched on.

\subsection{Time-dependent charge and energy currents}
As shown previously\cite{crsa09a}, the charge current flowing from reservoir $\alpha$
into the device is given by
\begin{equation}\label{eq:DefElCurr}
    J^{\rm C}_\alpha(t) = \frac{2 e}{\hbar}\Re\Tr \mbf{\Pi}_\alpha ( t )
        = \frac{2 e}{\hbar}\Re\Tr\left\{\frac{1}{4}\left( \mbf{1} - 2\mbf{\RDM}\right)\mbf{\Gamma}_\alpha + \sum_p \mbf{\Pi}_{\alpha p} (t) \right\}\;.
\end{equation}
Noting that the number of electrons in the device is given by $N_{\rm dev.} =\Tr \mbf{\RDM}$, one readily finds the continuity equation for the charge current
from Eq.\ \eqref{eq:EOMSEDM2},
\begin{equation}\label{eq:chargebalance}
    \dot{N}_{\rm dev.} = \Tr \dot{\mbf{\RDM}} = \Tr\sum_\alpha {2}\Re \mbf{\Pi}_\alpha = \sum_\alpha J^{\rm C}_\alpha/e\;.
\end{equation}
Note that the currents flowing out of the reservoirs are defined to be positive and thus increase the number of electrons
in the device.

Now, the electronic energy-current is given by an expression similar to Eq.\ \eqref{eq:DefElCurr}, namely
\begin{equation}\label{eq:DefECurr}
    J^{E}_\alpha(t) = {2}\Re\Tr \mbf{\Pi}^{E}_\alpha ( t ) \;,
\end{equation}
where new energy-current matrices are defined as
\begin{equation}
  \mbf{\Pi}^{E}_\alpha ( t ) = \int\limits^t_{t_0} dt_2
     \left( \mathbf{G}^>(t,t_2) \mbf{\tilde{\Sigma}}^<_\alpha (t_2,t)
       - \mathbf{G}^<(t,t_2) \mbf{\tilde{\Sigma}}^>_\alpha (t_2,t) \right) \;.
  \label{eq:DefPiE}
\end{equation}
The difference to the current matrices $\mbf{\Pi}_\alpha$ is the definition of the self-energies \cite{crsi+11,crsi+14}
\begin{align}
    \tilde{\bSigma}_\alpha^>(t_2,t) ={}& -\im \int \frac{\mathrm{d}\epsilon}{2\pi \hbar} \epsilon (1 - f_\alpha(\epsilon)) \bGamma_\alpha(\epsilon) e^{-i \epsilon (t_2-t)/\hbar}
    = -\im \partial_t \bSigma_\alpha^>(t_2,t) \;, \label{eq_lesser_1A}    \\
    \tilde{\bSigma}_\alpha^<(t_2,t) ={}& \im  \int \frac{\mathrm{d}\epsilon}{2\pi \hbar} \epsilon f_\alpha(\epsilon) \bGamma_\alpha(\epsilon) e^{-i \epsilon (t_2-t)/\hbar}
    = -\im \partial_t \bSigma_\alpha^<(t_2,t)\;,         \label{eq_lesser_1B}
\end{align}
which accounts for the energy transfer process. Since the energy is transported by the
electrons leaving and entering the device, the energy-current matrices can be related to
the charge-current matrices.
As shown in Appendix \ref{app:PiE}, the final expression for the energy current is
\begin{multline}\label{eq:ECurr}
    J^{E}_\alpha(t) = {2}\Re\Tr\left\{\left[
        -\mbf{\RDM} (t) \mbf{H} (t)
        + \mbf{H}/2  + \im \sum\limits_{\alpha' \in \text{leads}} \sum_p \mbf{\Pi}_{\alpha' p} (t) \right]\mbf{\Gamma}_\alpha/2 \right.\\
        + \left.\sum_p \chi^+_{\alpha p} \mbf{\Pi}_{\alpha p} (t)
        \right\} \;.
\end{multline}
Consequently, the energy current can be calculated once the auxiliary current matrices
are known. Therefore, no additional quantities have to be computed.
Similar to the charge-current continuity in Eq.\ \eqref{eq:chargebalance}, the energy currents can be interpreted by considering the variation of energy in the device, which leads to a corresponding continuity equation. The device energy is given
by $E_{\rm dev.} = \Tr\{\mbf{H}(t)\mbf{\RDM}\}$ and thus one finds
\begin{align}
    \frac{\partial}{\partial t}E_{\rm dev.}
    ={}& \frac{\partial}{\partial t}\Tr\{\mbf{H}(t)\mbf{\RDM}(t)\} \notag\\
    ={}& \Tr\{\dot{\mbf{H}}(t)\mbf{\RDM}(t)\}
    - \Tr\left\{\mbf{H}(t)\ACOMM{ \mbf{\Gamma}/2 }{ \mbf{\RDM}( t ) }\right\}\notag\\
     {}&+ \Tr\left\{\mbf{H}(t)\sum\limits_{\alpha \in \text{leads}} \left(   \frac{1}{4} \mbf{\Gamma}_\alpha + \sum_p \mbf{\Pi}_{\alpha p} (t) + h.c. \right)\right\} \notag\\
    ={}& \Tr\{\dot{\mbf{H}}(t)\mbf{\RDM}(t)\} \notag\\
     {}&+ \sum\limits_{\alpha \in \text{leads}} 2\Re\Tr\left\{-\frac{1}{2}\mbf{\RDM}( t )\mbf{H}(t) \mbf{\Gamma}_\alpha
     +\frac{1}{4} \mbf{H}(t) \mbf{\Gamma}_\alpha
     +\mbf{H}(t) \sum_p \mbf{\Pi}_{\alpha p} (t) \right\} \notag\\
    ={}& \Tr\{\dot{\mbf{H}}(t)\mbf{\RDM}(t)\} +
        \sum\limits_{\alpha \in \text{leads}} \left[ J^E_{\alpha}(t) + J^{E}_{c\alpha}(t) \right] \;. \label{eq:energybalance}
\end{align}
The last term gives an explicit expression for the energy variation of the
contact (coupling) between reservoirs and device,
\begin{equation}
    J^{E}_{c\alpha} (t) = \sum_p 2\Re\Tr\left\{
      \left(\mbf{H}(t) - \chi^+_{\alpha p}\mbf{1}\right)\mbf{\Pi}_{\alpha p} (t) - \frac{\im}{2} \sum\limits_{\alpha' \in \text{leads}} \mbf{\Pi}_{\alpha' p} (t) \mbf{\Gamma}_\alpha\right\}\;.
\end{equation}
As shown in Refs.\ \cite{luli+14,lumo+16} the contribution of the contacts to the energy balance is important to
obtain a consistent thermodynamic description.

Before coming to the results in the next section, it should be pointed out that in the WBL the energy current of each individual reservoir is infinite. This is due to the unboundedness of the electronic bands in the WBL. To obtain finite results one has to introduce a cutoff for the energy integrations. In the auxiliary-mode approach this cutoff is provided by the finite number of poles.
To get meaningful values for the energy current, we subtract the values obtained in the (initial) stationary state. In this way, the individual net energy-currents, $J^E_{\alpha}(t)-J^E_{\alpha}(t_0)$, yield finite values.

 \FloatBarrier
\section{Results}\label{sec:results}
We give two examples for the application of this approach: a single-level toy model and a benzene ring inspired by single-molecular junctions. In both cases, the system is connected to two reservoirs in thermal equilibrium. The coupling to the reservoirs is described by the level-width functions $\Gamma_{\rm L}=1,\ \Gamma_{\rm R}=0.5$, where we consider wide-band electrodes. All energies in this text are given in units of $\Gamma_{\rm L}$ and all time values in units of $\hbar/\Gamma_{\rm L}$. The respective chemical potentials were set to $\mu_{\rm L}=2,\ \mu_{\rm R}=1$ with an inverse temperature of $\beta=(k_{\rm B}T)^{-1}=20$. For the calculation we used $N_{\rm F}=40$ poles in the Fermi decomposition and a time-step of $dt=0.05$ with a 7th order Runge-Kutta solver. We compare the time-resolved currents with the corresponding stationary solutions, to which we refer to as \textit{instantaneous} solutions. Naturally, both are equal for the steady state and adiabatically slow time-dependencies.

\subsection{One-level model}
\begin{figure*}
  \center
  \includegraphics{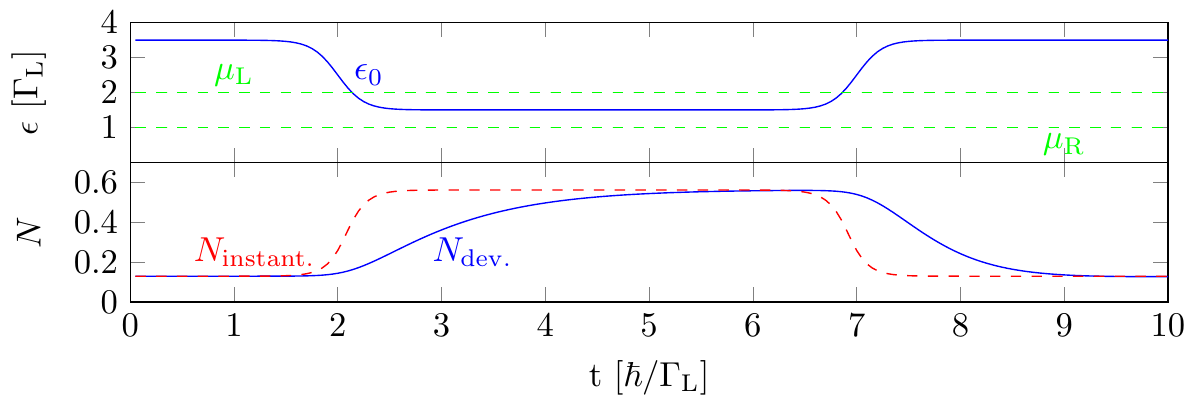}
  \caption{Time evolution of on-site energy $\epsilon_0(t)$ and occupation $N_{\rm {dev.}}(t)$ of the one-level model system. The chemical potentials $\mu_{\rm{L}},\ \mu_{\rm{R}}$ refer to the transport energy window. The instantaneous occupation $N_{\rm instant.}$ is obtained from the stationary solution of Eq.\ \eqref{eq:EOMSEDM2} for each $\epsilon_o(t)$.}
  \label{fig:OL1}
\end{figure*}

First, we consider a one-level device connected to source and drain electrodes. At time $t=2$ the on-site energy of the level is shifted from $\epsilon_0=3.5$ to $\epsilon_0=1.5$, i.e., into the center of the transport window $[\mu_{\rm L}, \mu_{\rm R}]$. This is reversed at time $t=7$. The time-dependence of the on-site energy is thus chosen to follow a double $\tanh(t)$, see Fig.\ \ref{fig:OL1}, describing a nearly rectangular pulse:

\begin{equation}
  \label{eq:td-onsite}
  \epsilon_0 = 3.5 + \tanh\left[ -2\cdot\frac{t-t_0}{t_w}\right] - \tanh\left[ -2\cdot\frac{t-(t_0+t_L)}{t_w}\right].
\end{equation}

\begin{figure*}
  \includegraphics{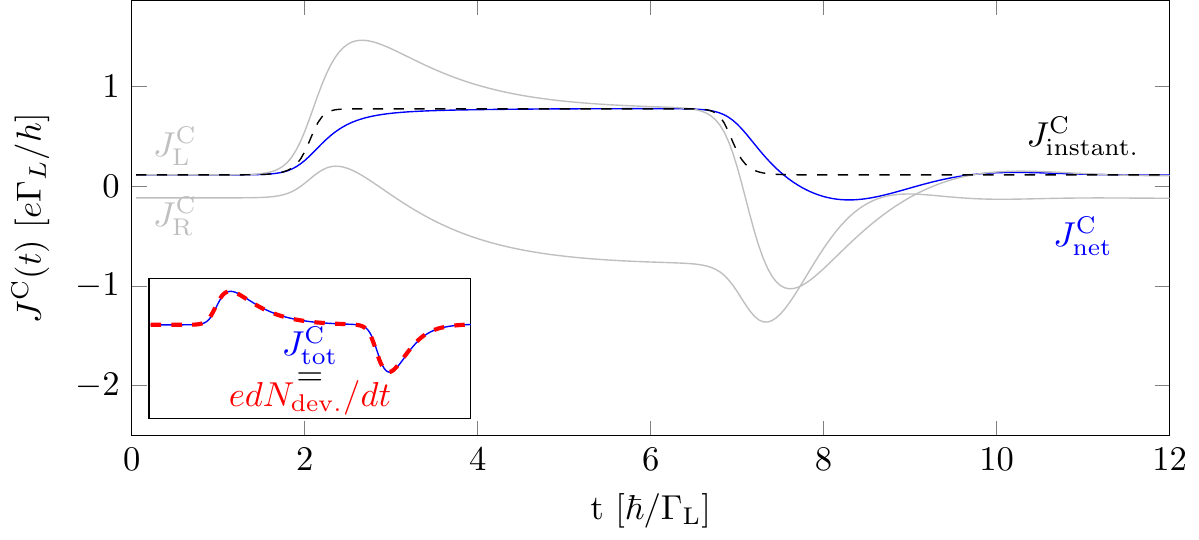}
  \includegraphics{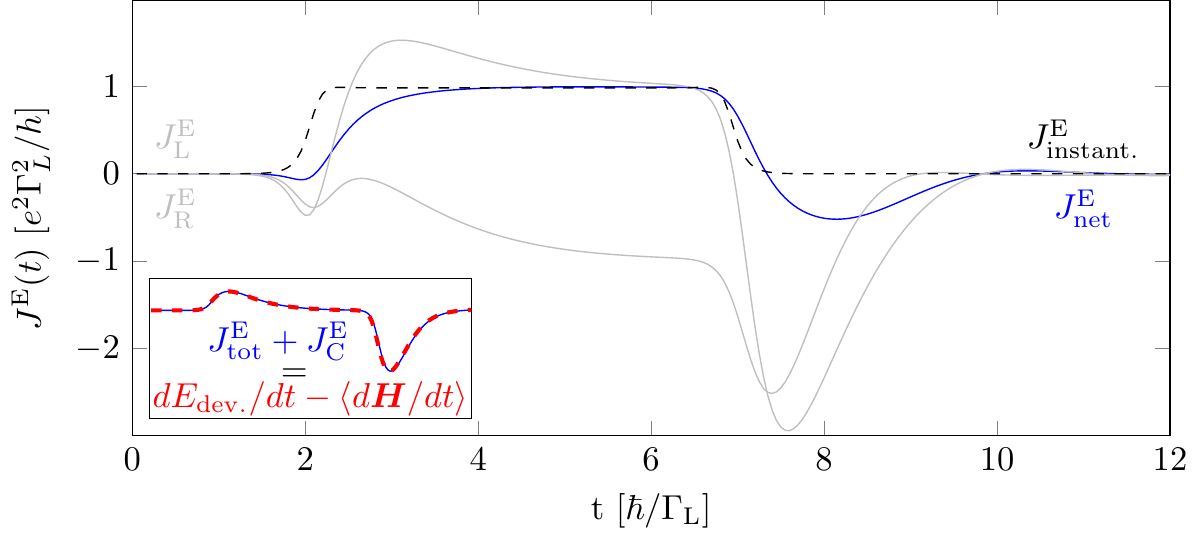}
  \caption{(Top) Time-dependent net charge current $(J^{\rm C}_{\rm L}-J^{\rm C}_{\rm R})/2$ and (bottom) net energy current $(J^{\rm E}_{\rm L}-J^{\rm E}_{\rm R})/2$ through the one-level system, as well as their individual components $J^{\rm C/E}_{\rm \alpha}$. The fullfillment of the corresponding conservation laws given by Eqs.\ \eqref{eq:chargebalance} and \eqref{eq:energybalance}, respectively, is shown in the insets.}
  \label{fig:OL2}
\end{figure*}

As it can be seen in Fig.\ \ref{fig:OL1}, the population and de-population of the energy level is retarded, compared to the instantaneous solution. This is also reflected in the charge and energy currents, see Fig.\ \ref{fig:OL2}, with overshoots appearing at the beginning and after the pulse, which eventually vanish. As expected, the time response of both currents is similar, since they are both driven by electrons. The insets in Fig.\ \ref{fig:OL2} demonstrate that the time-dependent solutions for the currents satisfy the corresponding conservation laws given by Eqs.\ \eqref{eq:chargebalance} and \eqref{eq:energybalance}, respectively. In the case of charge currents, the total current, $J^{\rm C}_{\rm L}+J^{\rm C}_{\rm R}$, corresponds to the change in occupation, $e dN_{\rm dev.}/dt$. For the energy current, the sum $J^E+J^E_c$ of the total energy current, $J^E = J^E_{\rm L}+J^E_{\rm R}$, and the interface current $J^E_{c{\rm L}}+J^E_{c{\rm R}}$, is equal to the change of energy in the system, given by $dE_{\rm dev.}/dt - \mean{d\mbf{H}/dt}$. The agreement with the conservation laws and the steady-state solutions validate our numerical approach.

\begin{figure*}
  \includegraphics{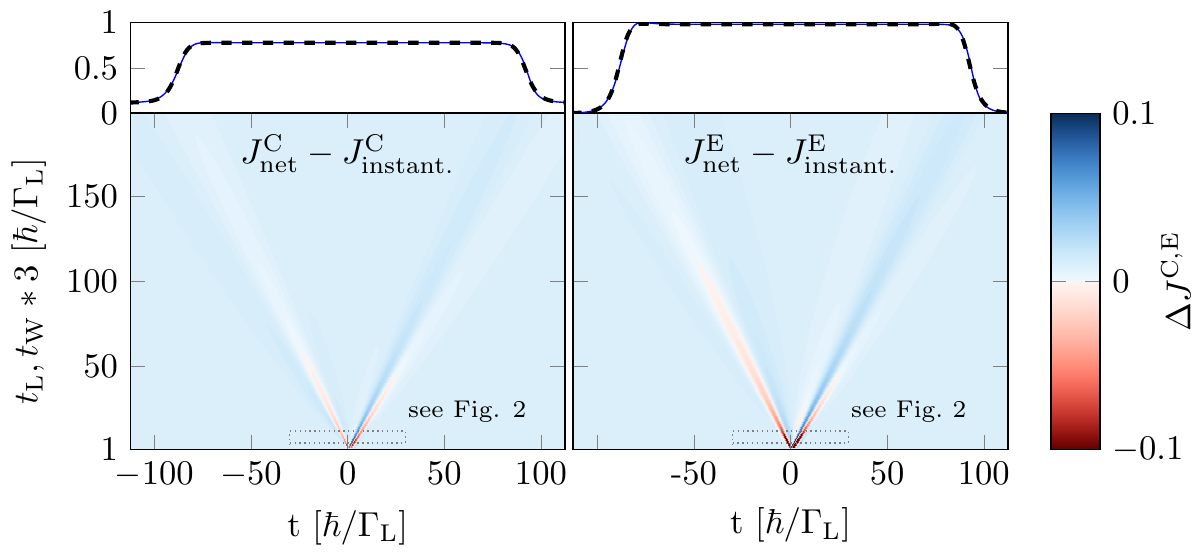}
  \caption{(Left) Time-dependent charge current $J^{\rm C}_{\rm net}$ and (right) energy current  $J^{\rm E}_{\rm net}$ in the one-level system for different pulse widths $t_{\rm W}$ and lengths $t_{\rm L}$, displayed as the difference between transient and instantaneous solution $\Delta J^{\rm C/E} = J^{\rm C/E}_{\rm {net}} - J^{\rm C/E}_{\rm {instant.}}$. This highlights the deviations from the adiabatic solution. The parameter range of  Fig.\ \ref{fig:OL2} is marked in the plot, additionally to the profile of the nearly adiabatic solution shown on top, which converges to the instantaneous solution $J^{\rm C/E}_{\rm {instant.}}$. The color bar indicates the difference in current renormalized respectively to a maximum absolute value of one.}
  \label{fig:OL3}
\end{figure*}

By changing the pulse, for example by considering different widths of the $\tanh(t)$ shape, one can demonstrate the transition from non-adiabatic to quasi-adiabatic behaviour. By smoothing out the step-edge in the time-domain, the transient solution converges to the instantaneous solution. In Fig.\ \ref{fig:OL3} we plot the difference of both solutions, i.e. transient solution minus instantaneous solution, $J^{\rm C/E}_{\rm {net}} - J^{\rm C/E}_{\rm {instant.}} $, over varying pulse lengths $t_{\rm L}$ and pulse widths $t_{\rm W}$. One can clearly identify the non-adiabatic contributions for short and sharp pulses, in contrast to the vanishing difference, thus quasi-adiabatic behaviour, for long and smooth pulses.

\subsection{Benzene}
To demonstrate that the presented approach is applicable to any multi-level system, we extend the one-level problem to a molecular junction, i.e., a benzene molecule contacted to two reservoirs. We chose a relatively small system for demonstration purpose, but this approach is notably not limited to small site numbers. The benzene molecule is modeled by a $6\times 6$ Hamiltonian in the standard H\"{u}ckel approach by considering only $p_z$ orbitals. For simplicity, the hopping elements are set to $V=-1$. In analogy to the one-level test system, all six on-site energies $\epsilon_n$ were shifted simultaneously from $\epsilon_n=3.5$ to $\epsilon_n=1.5$, following a $\tanh(t)$ dependence. Since the resulting transient is more complex than in the previous case, we do not consider the reverse shift out of the transport window, i.e., we only use the first half of the pulse from Eq.\ \eqref{eq:td-onsite}.

\begin{figure*}
\center
  \includegraphics{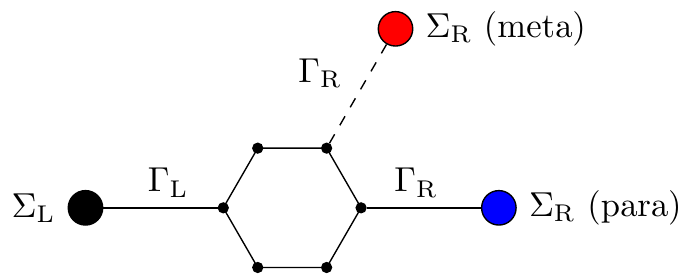}
  \caption{Schematic system setup of the molecular junction with different lead conficurations, called \textit{meta}-Benzene and \textit{para}-Benzene.}
  \label{fig:RB0}
\end{figure*}

\begin{figure*}
\center
  \includegraphics{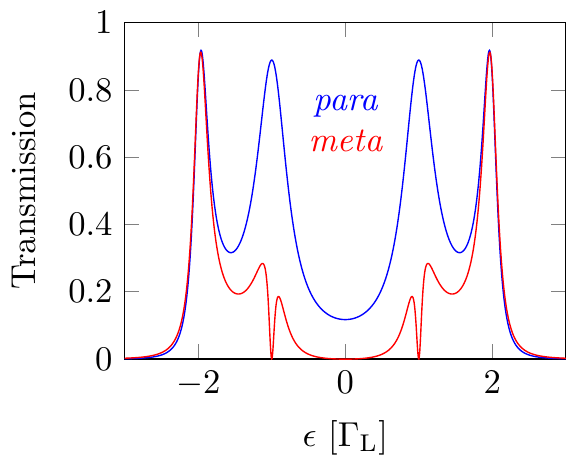}
  \caption{Transmission functions for \textit{meta}-Benzene and \textit{para}-Benzene over energy $\epsilon$, calculated by means of the standard (equilbrium) Green function formalism~\cite{Datta95book}.}
  \label{fig:RB1}
\end{figure*}

\begin{figure*}
  \includegraphics{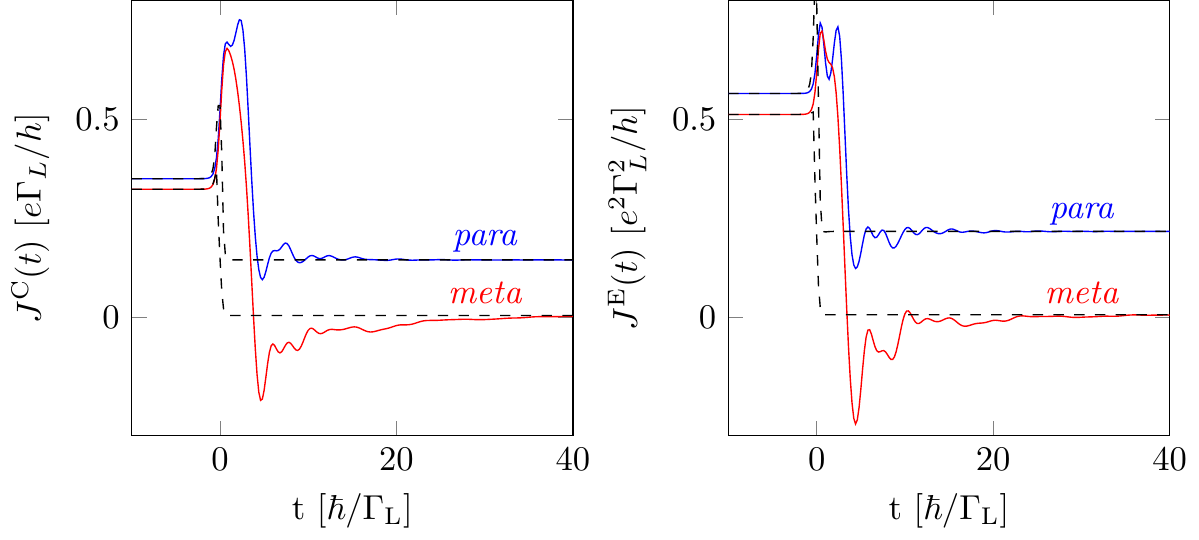}
  \caption{(Left) Time-dependent net charge current $J^{\rm C}_{\rm net}$ and (right) net energy current $J^{\rm E}_{\rm net}$ for \textit{meta}-Benzene and \textit{para}-Benzene in response of a $\tanh(t)$-pulse with a width of $t_{\rm W}=1\ \hbar/\Gamma_{\rm L}$. The energy currents are shifted by their stationary values for $t=-\infty$.}
  \label{fig:RB2}
\end{figure*}

\begin{figure*}
  \includegraphics{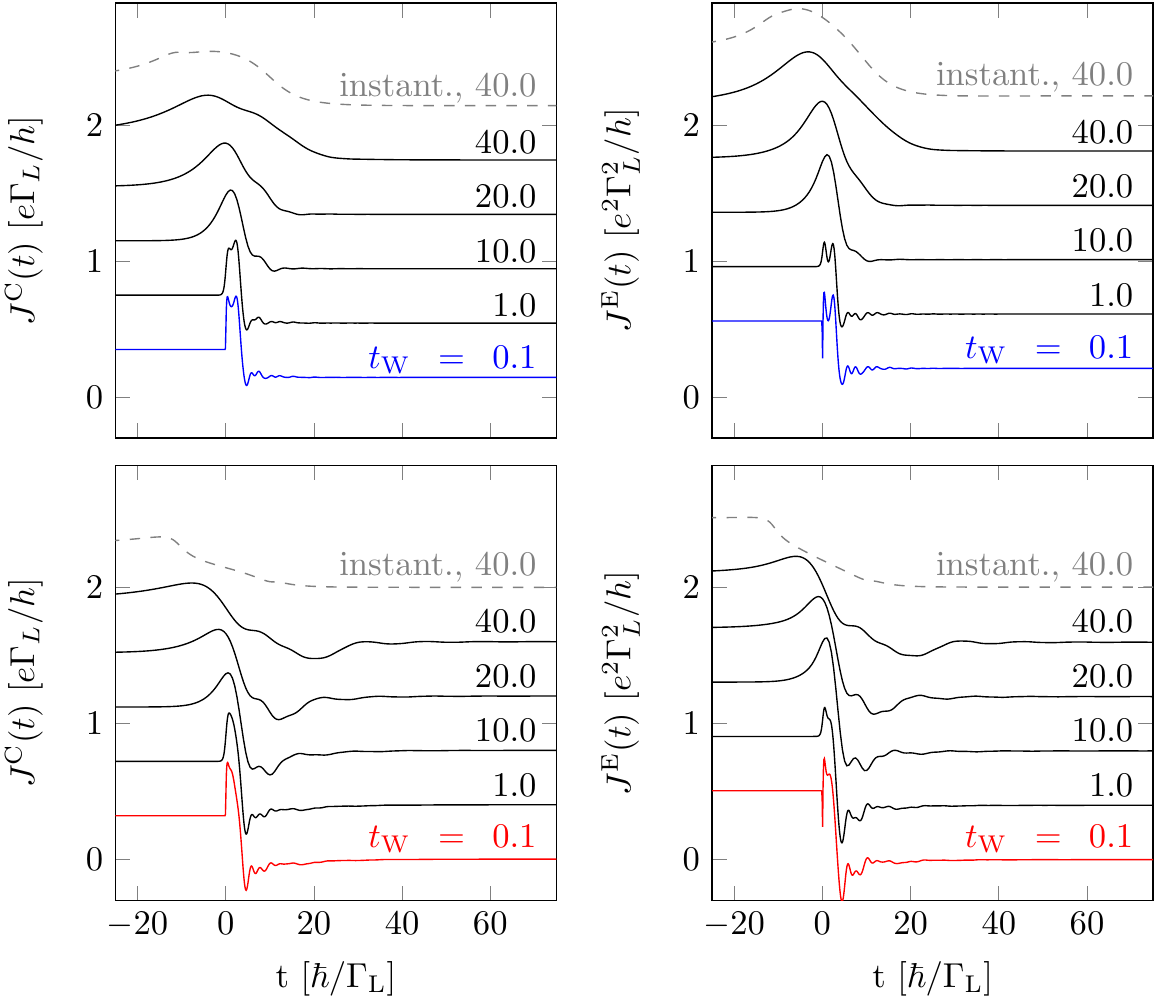}
   \caption{(Left) Time-dependent net charge current $J^{\rm C}_{\rm net}$ and (right) net energy current $J^{\rm E}_{\rm net}$ stacked for different pulse widths $t_{\rm W}$ for (top) \textit{para}-Benzene and (bottom) \textit{meta}-Benzene. The instantaneous solution is shown for comparison for the largest pulse-width (dashed line).}
  \label{fig:RB3}
\end{figure*}

One example for the use of time-resolved transport calculation is the dynamics of quantum interference. For illustration, we apply this approach on two different electrode configurations for that molecular junction: a symmetric and asymmetric setup, which we denote by \textit{para}-Benzene and \textit{meta}-Benzene respectively, see Fig.\ \ref{fig:RB0}. The electrons propagating through the ring utilize all possible paths between the electrodes. For \textit{meta}-Benzene, the most direct paths interfere destructively with a phase difference of $\pi$, which lowers the transmission function through the molecule, see Fig.\ \ref{fig:RB1}. The resulting current is thus lower than for the symmetric \textit{para}-Benzene setup and vanishes for $t=\infty$, where the transport window captures the center of the transmission function. In Fig.\ \ref{fig:RB2} we also see a significant difference in the transient, for charge and energy currents alike, before converging to the respective steady-state solutions. Most notably, we see a two-peak structure in the response of the \textit{para}-Benzene, especially for the energy current, which is a signature of the first two levels entering the transport window. For \textit{meta}-Benzene, the transmission for the inner two levels is suppressed by destructive interference, hence the single peak in the transient. The transient time is arguably the same for both cases. Figure\ \ref{fig:RB3} shows the same trend, i.e., by increasing the width of the  pulse a transition into quasi-adiabatic behaviour occurs.

 \section{Conclusion}\label{sec:summary}
In summary, we derived an extension of the auxiliary-mode approach for the description of the dynamics of nano-scale devices which allows the calculation of time-dependent energy currents. The latter can be calculated without introducing additional quantities. Although we focused specifically on the wide-band approximation, the idea presented in this article can also be applied to the more general case. From the energy-balance equation we identified an explicit expression for the contribution to the current due to the contacts, which is important for a thermodynamic description.

We applied the new scheme to two illustrative examples and demonstrated its validity by using the corresponding conservation laws and comparing to steady-state solutions. In both cases we find that charge and energy currents display a very similar transient behavior after switching, whereas the transient response of both currents is stronger for the \textit{meta}-configuration  compared to the \textit{para}-configuration. This effect might offer a path to tune charge and energy currents in systems showing quantum interference by optimizing pulse shapes.

Overall, our extension broadens the range of applications of the auxiliary-mode approach and offers a new tool to study the viability of molecular nanomachines.

 \clearpage

\section*{Acknowledgments}
This work is partly supported by the German Research Foundation (DFG) within the Cluster of Excellence "Center for Advancing Electronics Dresden" and project MEMO, funded by the European Union’s Horizon 2020 research and innovation programme under the grant agreement No. 766864. T.L. gratefully acknowledges the International Excellence Graduate School on Emerging Materials and Processes Korea (iEGSEMP Korea) in the context of TU Dresden’s Institutional Strategy The Synergetic University.
\section*{References}
\bibliography{levine}

\begin{thebibliography}{10}
\expandafter\ifx\csname url\endcsname\relax
  \def\url#1{\texttt{#1}}\fi
\expandafter\ifx\csname urlprefix\endcsname\relax\def\urlprefix{URL }\fi
\expandafter\ifx\csname href\endcsname\relax
  \def\href#1#2{#2} \def\path#1{#1}\fi

\bibitem{Remade2004}
F.~Remacle, I.~Willner, R.~D. Levine, {Nanowiring by molecules}, J. Phys. Chem.
  B 108~(47) (2004) 18129--18134.
\newblock \href {http://dx.doi.org/10.1021/jp047591q}
  {\path{doi:10.1021/jp047591q}}.

\bibitem{Remacle2006}
F.~Remacle, R.~D. Levine, \href{http://xlink.rsc.org/?DOI=b505696a}{{Electrical
  transport in saturated and conjugated molecular wires.}}, Faraday Discuss.
  131 (2006) 45--67; discussion 91--109.
\newblock \href {http://dx.doi.org/10.1039/b505696a}
  {\path{doi:10.1039/b505696a}}.
\newline\urlprefix\url{http://xlink.rsc.org/?DOI=b505696a}

\bibitem{Remacle2004}
F.~Remacle, R.~D. Levine, {Electronic and electrical response of arrays of
  metallic quantum dots}, Int. J. Quantum Chem. 99~(5 SPEC. ISS.) (2004)
  743--751.
\newblock \href {http://dx.doi.org/10.1002/qua.20047}
  {\path{doi:10.1002/qua.20047}}.

\bibitem{Mol2011}
J.~a. Mol, J.~Verduijn, R.~D. Levine, F.~Remacle, S.~Rogge, {Integrated logic
  circuits using single-atom transistors.}, Proc. Natl. Acad. Sci. U. S. A.
  108~(34) (2011) 13969--13972.
\newblock \href {http://dx.doi.org/10.1073/pnas.1109935108}
  {\path{doi:10.1073/pnas.1109935108}}.

\bibitem{doi:10.1063/1.4936182}
A.~Kleshchonok, R.~Gutierrez, C.~Joachim, G.~Cuniberti, Photoassisted transport
  in silicon dangling bond wires, Applied Physics Letters 107~(20) (2015)
  203109.
\newblock \href {http://dx.doi.org/10.1063/1.4936182}
  {\path{doi:10.1063/1.4936182}}.

\bibitem{Klymenko2016}
M.~V. Klymenko, M.~Klein, R.~D. Levine, F.~Remacle, {Operation of a quantum dot
  in the finite-state machine mode: Single-electron dynamic memory}, J. Appl.
  Phys. 120~(2) (2016) 024503.
\newblock \href {http://dx.doi.org/10.1063/1.4955422}
  {\path{doi:10.1063/1.4955422}}.

\bibitem{crsi+11}
A.~Cr{\'{e}}pieux, F.~{\v{S}}imkovic, B.~Cambon, F.~Michelini, {Enhanced
  thermopower under a time-dependent gate voltage}, Phys. Rev. B 83~(15) (2011)
  153417.
\newblock \href {http://dx.doi.org/10.1103/PhysRevB.83.153417}
  {\path{doi:10.1103/PhysRevB.83.153417}}.

\bibitem{crsi+14}
A.~Cr{\'{e}}pieux, F.~{\v{S}}imkovic, B.~Cambon, F.~Michelini, {Erratum:
  Enhanced thermopower under a time-dependent gate voltage [Phys. Rev. B 83 ,
  153417 (2011)]}, Phys. Rev. B 89~(23) (2014) 239907.
\newblock \href {http://dx.doi.org/10.1103/PhysRevB.89.239907}
  {\path{doi:10.1103/PhysRevB.89.239907}}.

\bibitem{dalo16}
A.-M. Dar{\'{e}}, P.~Lombardo, {Time-dependent thermoelectric transport for
  nanoscale thermal machines}, Phys. Rev. B 93~(3) (2016) 035303.
\newblock \href {http://dx.doi.org/10.1103/PhysRevB.93.035303}
  {\path{doi:10.1103/PhysRevB.93.035303}}.

\bibitem{luli+14}
M.~F. Ludovico, J.~S. Lim, M.~Moskalets, L.~Arrachea, D.~S{\'{a}}nchez,
  {Dynamical energy transfer in ac-driven quantum systems}, Phys. Rev. B
  89~(16) (2014) 161306.
\newblock \href {http://dx.doi.org/10.1103/PhysRevB.89.161306}
  {\path{doi:10.1103/PhysRevB.89.161306}}.

\bibitem{lumo+16}
M.~F. Ludovico, M.~Moskalets, D.~S{\'{a}}nchez, L.~Arrachea, {Dynamics of
  energy transport and entropy production in ac-driven quantum electron
  systems}, Phys. Rev. B 94~(3) (2016) 035436.
\newblock \href {http://dx.doi.org/10.1103/PhysRevB.94.035436}
  {\path{doi:10.1103/PhysRevB.94.035436}}.

\bibitem{kurt05}
S.~Kurth, G.~Stefanucci, C.-O. Almbladh, A.~Rubio, E.~K.~U. Gross,
  {Time-dependent quantum transport: A practical scheme using density
  functional theory}, Phys. Rev. B 72~(3) (2005) 035308.
\newblock \href {http://dx.doi.org/10.1103/PhysRevB.72.035308}
  {\path{doi:10.1103/PhysRevB.72.035308}}.

\bibitem{zhu05a}
Y.~Zhu, J.~Maciejko, T.~Ji, H.~Guo, J.~Wang, {Time-dependent quantum transport:
  Direct analysis in the time domain}, Phys. Rev. B 71~(7) (2005) 075317.
\newblock \href {http://dx.doi.org/10.1103/PhysRevB.71.075317}
  {\path{doi:10.1103/PhysRevB.71.075317}}.

\bibitem{mold07a}
V.~Moldoveanu, V.~Gudmundsson, A.~Manolescu, {Transient regime in nonlinear
  transport through many-level quantum dots}, Phys. Rev. B 76~(8) (2007)
  085330.
\newblock \href {http://dx.doi.org/10.1103/PhysRevB.76.085330}
  {\path{doi:10.1103/PhysRevB.76.085330}}.

\bibitem{prok08a}
A.~Prociuk, B.~D. Dunietz, {Modeling time-dependent current through electronic
  open channels using a mixed time-frequency solution to the electronic
  equations of motion}, Phys. Rev. B 78~(16) (2008) 165112.
\newblock \href {http://dx.doi.org/10.1103/PhysRevB.78.165112}
  {\path{doi:10.1103/PhysRevB.78.165112}}.

\bibitem{croy09a}
A.~Croy, U.~Saalmann, {Propagation scheme for nonequilibrium dynamics of
  electron transport in nanoscale devices}, Phys. Rev. B 80~(24) (2009) 245311.
\newblock \href {http://dx.doi.org/10.1103/PhysRevB.80.245311}
  {\path{doi:10.1103/PhysRevB.80.245311}}.

\bibitem{gaur14}
B.~Gaury, J.~Weston, M.~Santin, M.~Houzet, C.~Groth, X.~Waintal, {Numerical
  simulations of time-resolved quantum electronics}, Phys. Rep. 534~(1) (2014)
  1--37.
\newblock \href {http://dx.doi.org/10.1016/j.physrep.2013.09.001}
  {\path{doi:10.1016/j.physrep.2013.09.001}}.

\bibitem{ridl15a}
M.~Ridley, A.~MacKinnon, L.~Kantorovich, {Calculation of the current response
  in a nanojunction for an arbitrary time-dependent bias: application to the
  molecular wire}, J. Phys. Conf. Ser. 696 (2016) 012017.
\newblock \href {http://dx.doi.org/10.1088/1742-6596/696/1/012017}
  {\path{doi:10.1088/1742-6596/696/1/012017}}.

\bibitem{wing93}
N.~S. Wingreen, A.-P. Jauho, Y.~Meir, {Time-dependent transport through a
  mesoscopic structure}, Phys. Rev. B 48~(11) (1993) 8487--8490.
\newblock \href {http://dx.doi.org/10.1103/PhysRevB.48.8487}
  {\path{doi:10.1103/PhysRevB.48.8487}}.

\bibitem{jauh94}
A.-P. Jauho, N.~S. Wingreen, Y.~Meir, {Time-dependent transport in interacting
  and noninteracting resonant-tunneling systems}, Phys. Rev. B 50~(8) (1994)
  5528--5544.
\newblock \href {http://dx.doi.org/10.1103/PhysRevB.50.5528}
  {\path{doi:10.1103/PhysRevB.50.5528}}.

\bibitem{haug96}
H.~Haug, A.-P. Jauho, {Quantum Kinetics in Transport and Optics of
  Semiconductors}, Vol. 123 of Solid-State Sciences, Springer Berlin
  Heidelberg, Berlin, Heidelberg, 2008.
\newblock \href {http://dx.doi.org/10.1007/978-3-540-73564-9}
  {\path{doi:10.1007/978-3-540-73564-9}}.

\bibitem{crsa09a}
A.~Croy, U.~Saalmann, {Propagation scheme for nonequilibrium dynamics of
  electron transport in nanoscale devices}, Phys. Rev. B 80~(24) (2009) 245311.
\newblock \href {http://dx.doi.org/10.1103/PhysRevB.80.245311}
  {\path{doi:10.1103/PhysRevB.80.245311}}.

\bibitem{pocr16}
B.~S. Popescu, A.~Croy, {Efficient auxiliary-mode approach for time-dependent
  nanoelectronics}, New J. Phys. 18~(9) (2016) 093044.
\newblock \href {http://dx.doi.org/10.1088/1367-2630/18/9/093044}
  {\path{doi:10.1088/1367-2630/18/9/093044}}.

\bibitem{croy12a}
A.~Croy, U.~Saalmann, A.~R. Hern{\'{a}}ndez, C.~H. Lewenkopf, {Nonadiabatic
  electron pumping through interacting quantum dots}, Phys. Rev. B 85~(3)
  (2012) 035309.
\newblock \href {http://dx.doi.org/10.1103/PhysRevB.85.035309}
  {\path{doi:10.1103/PhysRevB.85.035309}}.

\bibitem{croy12b}
A.~Croy, U.~Saalmann, {Nonadiabatic rectification and current reversal in
  electron pumps}, Phys. Rev. B 86~(3) (2012) 035330.
\newblock \href {http://dx.doi.org/10.1103/PhysRevB.86.035330}
  {\path{doi:10.1103/PhysRevB.86.035330}}.

\bibitem{pope12a}
B.~Popescu, P.~B. Woiczikowski, M.~Elstner, U.~Kleinekath{\"{o}}fer,
  {Time-Dependent View of Sequential Transport through Molecules with Rapidly
  Fluctuating Bridges}, Phys. Rev. Lett. 109~(17) (2012) 176802.
\newblock \href {http://dx.doi.org/10.1103/PhysRevLett.109.176802}
  {\path{doi:10.1103/PhysRevLett.109.176802}}.

\bibitem{chen14c}
S.~Chen, Y.~Zhang, S.~Koo, H.~Tian, C.~Yam, G.~Chen, M.~A. Ratner,
  {Interference and Molecular Transport a Dynamical View: Time-Dependent
  Analysis of Disubstituted Benzenes}, J. Phys. Chem. Lett. 5~(15) (2014)
  2748--2752.
\newblock \href {http://dx.doi.org/10.1021/jz5007143}
  {\path{doi:10.1021/jz5007143}}.

\bibitem{xie13a}
H.~Xie, Y.~Kwok, Y.~Zhang, F.~Jiang, X.~Zheng, Y.~Yan, G.~Chen, {Time-dependent
  quantum transport theory and its applications to graphene nanoribbons}, Phys.
  status solidi 250~(11) (2013) 2481--2494.
\newblock \href {http://dx.doi.org/10.1002/pssb.201349247}
  {\path{doi:10.1002/pssb.201349247}}.

\bibitem{cao15a}
H.~Cao, M.~Zhang, T.~Tao, M.~Song, C.~Zhang, {Electric response of a
  metal-molecule-metal junction to laser pulse by solving hierarchical
  equations of motion}, J. Chem. Phys. 142~(8) (2015) 084705.
\newblock \href {http://dx.doi.org/10.1063/1.4913466}
  {\path{doi:10.1063/1.4913466}}.

\bibitem{wang13a}
R.~Wang, D.~Hou, X.~Zheng, {Time-dependent density-functional theory for
  real-time electronic dynamics on material surfaces}, Phys. Rev. B 88~(20)
  (2013) 205126.
\newblock \href {http://dx.doi.org/10.1103/PhysRevB.88.205126}
  {\path{doi:10.1103/PhysRevB.88.205126}}.

\bibitem{pocr17}
B.~S. Popescu, A.~Croy, {Emergence of Bloch oscillations in one-dimensional
  systems}, Phys. Rev. B 95~(23) (2017) 235433.
\newblock \href {http://dx.doi.org/10.1103/PhysRevB.95.235433}
  {\path{doi:10.1103/PhysRevB.95.235433}}.

\bibitem{kuel+17}
T.~Kubar, M.~Elstner, B.~Popescu, U.~Kleinekath{\"{o}}fer, {Polaron Effects on
  Charge Transport through Molecular Wires: A Multiscale Approach}, J. Chem.
  Theory Comput. 13~(1) (2017) 286--296.
\newblock \href {http://dx.doi.org/10.1021/acs.jctc.6b00879}
  {\path{doi:10.1021/acs.jctc.6b00879}}.

\bibitem{PhysRevLett.102.208102}
R.~Guti\'errez, R.~A. Caetano, B.~P. Woiczikowski, T.~Kubar, M.~Elstner,
  G.~Cuniberti, Charge transport through biomolecular wires in a solvent:
  Bridging molecular dynamics and model hamiltonian approaches, Phys. Rev.
  Lett. 102 (2009) 208102.
\newblock \href {http://dx.doi.org/10.1103/PhysRevLett.102.208102}
  {\path{doi:10.1103/PhysRevLett.102.208102}}.

\bibitem{PhysRevB.82.155455}
M.~H. Lee, S.~Avdoshenko, R.~Gutierrez, G.~Cuniberti, Charge migration through
  dna molecules in the presence of mismatches, Phys. Rev. B 82 (2010) 155455.
\newblock \href {http://dx.doi.org/10.1103/PhysRevB.82.155455}
  {\path{doi:10.1103/PhysRevB.82.155455}}.

\bibitem{doi:10.1021/ja078162j}
F.~C. Grozema, S.~Tonzani, Y.~A. Berlin, G.~C. Schatz, L.~D.~A. Siebbeles,
  M.~A. Ratner, {Effect of Structural Dynamics on Charge Transfer in DNA
  Hairpins}, J. Am. Chem. Soc. 130~(15) (2008) 5157--5166.
\newblock \href {http://dx.doi.org/10.1021/ja078162j}
  {\path{doi:10.1021/ja078162j}}.

\bibitem{doi:10.1021/jp004436c}
Y.~A. Berlin, A.~L. Burin, L.~D.~A. Siebbeles, M.~A. Ratner, {Conformationally
  Gated Rate Processes in Biological Macromolecules †}, J. Phys. Chem. A
  105~(23) (2001) 5666--5678.
\newblock \href {http://dx.doi.org/10.1021/jp004436c}
  {\path{doi:10.1021/jp004436c}}.

\bibitem{ANIE:ANIE201207667}
C.~R. Arroyo, S.~Tarkuc, R.~Frisenda, J.~S. Seldenthuis, C.~H.~M. Woerde,
  R.~Eelkema, F.~C. Grozema, H.~S.~J. van~der Zant, Signatures of quantum
  interference effects on charge transport through a single benzene ring,
  Angewandte Chemie International Edition 52~(11) (2013) 3152--3155.

\bibitem{PhysRevB.59.16011}
M.~Magoga, C.~Joachim,
  \href{https://link.aps.org/doi/10.1103/PhysRevB.59.16011}{Conductance of
  molecular wires connected or bonded in parallel}, Phys. Rev. B 59 (1999)
  16011--16021.
\newblock \href {http://dx.doi.org/10.1103/PhysRevB.59.16011}
  {\path{doi:10.1103/PhysRevB.59.16011}}.
\newline\urlprefix\url{https://link.aps.org/doi/10.1103/PhysRevB.59.16011}

\bibitem{doi:10.1021/acs.jpclett.5b01595}
D.~Nozaki, Lokamani, A.~Santana-Bonilla, A.~Dianat, R.~Gutierrez, G.~Cuniberti,
  \href{http://dx.doi.org/10.1021/acs.jpclett.5b01595}{Switchable negative
  differential resistance induced by quantum interference effects in
  porphyrin-based molecular junctions}, The Journal of Physical Chemistry
  Letters 6~(19) (2015) 3950--3955, pMID: 26722897.
\newblock \href
  {http://arxiv.org/abs/http://dx.doi.org/10.1021/acs.jpclett.5b01595}
  {\path{arXiv:http://dx.doi.org/10.1021/acs.jpclett.5b01595}}, \href
  {http://dx.doi.org/10.1021/acs.jpclett.5b01595}
  {\path{doi:10.1021/acs.jpclett.5b01595}}.
\newline\urlprefix\url{http://dx.doi.org/10.1021/acs.jpclett.5b01595}

\bibitem{doi:10.1021/acs.jpclett.6b02989}
D.~Nozaki, A.~L{\"{u}}cke, W.~G. Schmidt, {Molecular Orbital Rule for Quantum
  Interference in Weakly Coupled Dimers: Low-Energy Giant Conductivity
  Switching Induced by Orbital Level Crossing}, J. Phys. Chem. Lett. 8~(4)
  (2017) 727--732.
\newblock \href {http://dx.doi.org/10.1021/acs.jpclett.6b02989}
  {\path{doi:10.1021/acs.jpclett.6b02989}}.

\bibitem{doi:10.1021/nl101688a}
T.~Markussen, R.~Stadler, K.~S. Thygesen, {The Relation between Structure and
  Quantum Interference in Single Molecule Junctions}, Nano Lett. 10~(10) (2010)
  4260--4265.
\newblock \href {http://dx.doi.org/10.1021/nl101688a}
  {\path{doi:10.1021/nl101688a}}.

\bibitem{doi:10.1063/1.2958275}
G.~C. Solomon, D.~Q. Andrews, T.~Hansen, R.~H. Goldsmith, M.~R. Wasielewski,
  R.~P. {Van Duyne}, M.~A. Ratner, {Understanding quantum interference in
  coherent molecular conduction}, J. Chem. Phys. 129~(5) (2008) 054701.
\newblock \href {http://dx.doi.org/10.1063/1.2958275}
  {\path{doi:10.1063/1.2958275}}.

\bibitem{doi:10.1021/jp9117216}
A.~A. Kocherzhenko, F.~C. Grozema, L.~D.~A. Siebbeles, {Charge Transfer Through
  Molecules with Multiple Pathways: Quantum Interference and Dephasing}, J.
  Phys. Chem. C 114~(17) (2010) 7973--7979.
\newblock \href {http://dx.doi.org/10.1021/jp9117216}
  {\path{doi:10.1021/jp9117216}}.

\bibitem{zhwa+07}
X.~Zheng, F.~Wang, C.~Y. Yam, Y.~Mo, G.~Chen, Time-dependent density-functional
  theory for open systems, Phys. Rev. B 75 (2007) 195127.
\newblock \href {http://dx.doi.org/10.1103/PhysRevB.75.195127}
  {\path{doi:10.1103/PhysRevB.75.195127}}.

\bibitem{oz07}
T.~Ozaki, {Continued fraction representation of the Fermi-Dirac function for
  large-scale electronic structure calculations}, Phys. Rev. B 75~(3) (2007)
  035123.
\newblock \href {http://dx.doi.org/10.1103/PhysRevB.75.035123}
  {\path{doi:10.1103/PhysRevB.75.035123}}.

\bibitem{kame+10}
C.~Karrasch, V.~Meden, K.~Sch{\"{o}}nhammer, {Finite-temperature linear
  conductance from the Matsubara Green's function without analytic continuation
  to the real axis}, Phys. Rev. B 82~(12) (2010) 125114.
\newblock \href {http://dx.doi.org/10.1103/PhysRevB.82.125114}
  {\path{doi:10.1103/PhysRevB.82.125114}}.

\bibitem{Datta95book}
S.~Datta, {Electronic Transport in Mesoscopic Systems}, Cambridge University
  Press, Cambridge, 1995.

\end{thebibliography}
\clearpage
\appendix
\section{Derivation of energy current matrix}\label{app:PiE}
The equations of motion for the greater and lesser Green functions $\mbf{G}^\gtrless$ in WBL are given by
\begin{equation}\label{eq:GglEOM}
    \im \partial_t \mbf{G}^\gtrless (t,t') = (\mbf{H} -\im \bGamma/2)\mbf{G}^\gtrless (t,t')
        + \int dt_1 \bSigma^\gtrless(t,t_1) \mbf{G}^{\rm a}(t_1, t')\;.
\end{equation}
Further, we have $\mbf{G}^<(t,t) = \im \mbf{\RDM}$ and $\mbf{G}^>(t,t) = -\im (\mbf{1}-\mbf{\RDM})$.

Using the definiton of the energy-current matrix, Eq.\ \eqref{eq:DefPiE}, and the expressions for the self-energies, Eqs.\ \eqref{eq_lesser_1A} and \eqref{eq_lesser_1B}, one finds for $\mbf{\Pi}^{E}_\alpha$
\begin{align}
  \mbf{\Pi}^{E}_\alpha ( t ) ={}& \frac{1}{\im} \partial_t \mbf{\Pi}_\alpha
    - \frac{1}{\im}\int\limits^t_{t_0} dt_2
     \left[   \partial_t\mathbf{G}^>(t,t_2) \mbf{{\Sigma}}^<_\alpha (t_2,t)
            - \partial_t\mathbf{G}^<(t,t_2) \mbf{{\Sigma}}^>_\alpha (t_2,t)
    \right]\notag\\
 {}&+ \im\left[ \mathbf{G}^>(t,t) \mbf{{\Sigma}}^<_\alpha (t,t)
              - \mathbf{G}^<(t,t) \mbf{{\Sigma}}^>_\alpha (t,t) \right]\;.
\end{align}
Here, we used the product rule to move the derivate from the self-energy
to the Green functions. Next we use the equations of motion \eqref{eq:GglEOM},
\eqref{eq:EOMSEDM2} and \eqref{eq:EOMPi} to further simplify the expression. One finds
\begin{align}
  \mbf{\Pi}^{E}_\alpha ( t )
={}& -\im \partial_t\frac{1}{4}\left( \mbf{1} - 2\mbf{\RDM}\right)\mbf{\Gamma}_\alpha -\im \partial_t \sum_p \mbf{\Pi}_{\alpha p} (t)
    + (\mbf{H} -\im \bGamma/2) \mbf{\Pi}_\alpha \notag\\
{}&+ \im\left( \mathbf{G}^>(t,t) \mbf{{\Sigma}}^<_\alpha (t,t)
    - \mathbf{G}^<(t,t) \mbf{{\Sigma}}^>_\alpha (t,t) \right) \notag\\
&   + \int\limits^t_{t_0} dt_2 \int dt_1
     \left( \bSigma^>(t,t_1) \mbf{G}^{\rm a}(t_1, t_2) \mbf{{\Sigma}}^<_\alpha (t_2,t)
    - \bSigma^>(t,t_1) \mbf{G}^{\rm a}(t_1, t_2) \mbf{{\Sigma}}^>_\alpha (t_2,t) \right) \notag\\
={}& \left\{ -\mbf{\RDM}(\mbf{H} +\im \bGamma/2)
        + (\mbf{H} +\im \bGamma/2)/2 + \im \sum\limits_{\alpha' \in \text{leads}} \left( \sum_p \mbf{\Pi}_{\alpha' p} (t) + \mbf{\Pi}_{\alpha' p}^\dagger (t) \right)\right\}\mbf{\Gamma}_\alpha/2 \notag\\
      & + \sum_p \left\{ \chi^+_{\alpha p} \mbf{\Pi}_{\alpha p}\right\}
        + \im\left( \mbf{1}/2
        - \mbf{\RDM}(t) \right)\bGamma_\alpha \delta(0) \notag\\
      & + \int\limits^t_{t_0} dt_2 \int dt_1
           \left( \bSigma^>(t,t_1) \mbf{G}^{\rm a}(t_1, t_2) \mbf{{\Sigma}}^<_\alpha (t_2,t)
        - \bSigma^<(t,t_1) \mbf{G}^{\rm a}(t_1, t_2) \mbf{{\Sigma}}^>_\alpha (t_2,t) \right)        \;.
\end{align}
It now remains to simplify the expression in the last line. Using
\begin{align}
  \mbf{\Sigma}^{\gtrless}_\alpha (t_1,t)
  ={}& \mp\im \frac{1}{2} \mbf{\Gamma}_\alpha \delta(t-t_1)
    + \sum\limits_p \mbf{\Sigma}_{\alpha p} (t_1,t) \;, \\
  \mbf{\Sigma}_{\alpha p} (t_1,t) ={}&  \frac{R_p}{\beta} \mbf{\Gamma}_\alpha
        e^{\im \int^t_{t_1} dt_2 \chi^+_{\alpha p} (t_2)}\;,
\end{align}
one finds
\begin{align}
    & \int\limits^t_{t_0} dt_2 \int dt_1
         \left( \bSigma^>(t,t_1) \mbf{G}^{\rm a}(t_1, t_2) \mbf{{\Sigma}}^<_\alpha (t_2,t)
      - \bSigma^<(t,t_1) \mbf{G}^{\rm a}(t_1, t_2) \mbf{{\Sigma}}^>_\alpha (t_2,t) \right) \notag\\
    ={}& \int\limits^t_{t_0} dt_2 \int dt_1
         \left( \bSigma^>(t,t_1) \mbf{G}^{\rm a}(t_1, t_2) \im \frac{1}{2} \mbf{\Gamma}_\alpha \delta(t-t_2)
       + \bSigma^<(t,t_1) \mbf{G}^{\rm a}(t_1, t_2) \im \frac{1}{2} \mbf{\Gamma}_\alpha \delta(t-t_1) \right)\notag\\
    &  + \sum_p\int\limits^t_{t_0} dt_2 \int dt_1
           \left( \bSigma^>(t,t_1) \mbf{G}^{\rm a}(t_1, t_2) \mbf{\Sigma}_{\alpha p} (t_2,t)
        - \bSigma^<(t,t_1) \mbf{G}^{\rm a}(t_1, t_2) \mbf{\Sigma}_{\alpha p} (t_2,t) \right)\notag\\
    ={}& \im \frac{1}{4} \int dt_1
         \left( \bSigma^>(t,t_1) + \bSigma^<(t,t_1)  \right)\mbf{G}^{\rm a}(t_1, t)\mbf{\Gamma}_\alpha\notag\\
    &  + \sum_p\int\limits^t_{t_0} dt_2 \int dt_1
           \left( \bSigma^>(t,t_1)
        - \bSigma^<(t,t_1)  \right) \mbf{G}^{\rm a}(t_1, t_2) \mbf{\Sigma}_{\alpha p} (t_2,t)\notag\\
    ={}& \im \frac{1}{2} \sum_{\alpha', p}\int dt_1
          \bSigma_{\alpha' p}(t,t_1)  \mbf{G}^{\rm a}(t_1, t)\mbf{\Gamma}_\alpha\notag\\
    &   -\im\sum_p\int\limits^t_{t_0} dt_2 \int dt_1
            \bGamma \delta(t-t_1)
         \mbf{G}^{\rm a}(t_1, t_2) \mbf{\Sigma}_{\alpha p} (t_2,t)\notag\\
    ={}& -\im \frac{1}{2} \sum_{\alpha', p} \mbf{\Pi}_{\alpha' p}^\dagger \mbf{\Gamma}_\alpha\;.
\end{align}
In the last step the second term vanishes since $t_2\leq t$ and $t_1\leq t_2$.

Finally, one obtains the expression for the energy-current from Eq.\ \eqref{eq:DefECurr}
\begin{equation}\label{eq:AppECurr}
    J^{E}_\alpha(t) = {2}\Re\Tr\left\{\left[
        -\mbf{\RDM} \mbf{H}
        + \mbf{H}/2  + \im \sum\limits_{\alpha' \in \text{leads}} \sum_p \mbf{\Pi}_{\alpha' p} (t) \right]\mbf{\Gamma}_\alpha/2 \notag\\
        + \sum_p \chi^+_{\alpha p} \mbf{\Pi}_{\alpha p}
        \right\} \;.
\end{equation}
Some of the terms vanished due to the hermiticity of the matrices and the
properties of the trace. Equation \eqref{eq:AppECurr} is the final result,
given as Eq.\ \eqref{eq:ECurr} in the main text.

\end{document}